\documentclass{article}
\usepackage{spconf,amsmath,graphicx,hyperref, xcolor, algorithm}
\usepackage{amsmath}
\usepackage{array}
\usepackage{tabularx}
\usepackage{booktabs}
\usepackage{color}
\usepackage{subfigure,amsmath,amssymb,amsfonts,bm,epsfig,epsf,url,dsfont}
\usepackage{amsthm,mathrsfs}
\usepackage{comment}

\usepackage{caption}
\usepackage{setspace}
\usepackage{amsmath}

\makeatletter
\g@addto@macro\normalsize{%
  \setlength{\abovedisplayskip}{1pt}
  \setlength{\belowdisplayskip}{1pt}
  \setlength{\abovedisplayshortskip}{0pt}
  \setlength{\belowdisplayshortskip}{0pt}
}
\makeatother

\DeclareMathOperator*{\argmin}{arg\,min}

\usepackage{algpseudocode}
\algrenewcommand\algorithmicrequire{\textbf{Input:}}
\algrenewcommand\algorithmicensure{\textbf{Output:}}

\usepackage{mathtools}

\title{SOLVING ILL-CONDITIONED POLYNOMIAL EQUATIONS USING SCORE-BASED PRIORS WITH APPLICATION TO MULTI-TARGET DETECTION}
\name{Rafi Beinhorn, Shay Kreymer, Amnon Balanov,  Michael Cohen, Alon Zabatani, and Tamir Bendory\thanks{T.B. is supported in part by BSF under Grant 2020159, in part by NSF-BSF under Grant 2024791, in part by ISF under Grant 1924/21, and in part by a grant from The Center for AI and Data Science at Tel Aviv University.}}
\address{School of Electrical and Computer Engineering, Tel Aviv University, Tel Aviv, Israel}

\let\oldbibliography\thebibliography
\renewcommand{\thebibliography}[1]{%
  \oldbibliography{#1}%
  \setlength{\itemsep}{-2.5pt}%
}

\begin{document}
%
\maketitle
\begin{abstract}
Recovering signals from low-order moments is a fundamental yet notoriously difficult task in inverse problems. This recovery process often reduces to solving ill-conditioned systems of polynomial equations. In this work, we propose a new framework that integrates score-based diffusion priors with moment-based estimators to regularize and solve these nonlinear inverse problems. This introduces a new role for generative models: stabilizing polynomial recovery from noisy statistical features. As a concrete application, we study the multi-target detection (MTD) model in the high-noise regime. We demonstrate two main results: (i) diffusion priors substantially improve recovery from third-order moments, and (ii) they make the super-resolution MTD problem, otherwise ill-posed, feasible. Numerical experiments on MNIST data confirm consistent gains in reconstruction accuracy across SNR levels. Our results suggest a promising new direction for combining generative priors with nonlinear polynomial inverse problems.
\end{abstract}

\section{Introduction}
The reconstruction of signals from limited and noisy measurements is a fundamental challenge in imaging, communications, and machine learning. A particularly difficult case arises when one aims to \emph{infer signals from low-order moments}, leading to polynomial systems that are typically ill-conditioned or even underdetermined~\cite{lasserre2001global, parrilo2003semidefinite}.
In some settings, such as super-resolution, the problem is fundamentally ill-posed without strong prior knowledge~\cite{donoho1992superresolution,  candes2014towards}.

Recovery from low-order moments can be formulated as solving a system of polynomial equations. The hardness of polynomial recovery is well known: solving general polynomial systems is NP-hard~\cite{parrilo2003semidefinite}. Classical approaches, including tensor decompositions for latent variable models~\cite{anandkumar2014tensor} and convex relaxations based on sum-of-squares, semidefinite programming, or the Lasserre hierarchy~\cite{lasserre2001global, parrilo2003semidefinite}, offer theoretical guarantees under restrictive assumptions but are computationally intensive and often unstable at low SNR. Recent work therefore seeks more scalable alternatives~\cite{rout2023solving, chen2024ntire}.

\begin{figure}[!tbp]
    \centering
    \includegraphics[width=0.9\columnwidth]{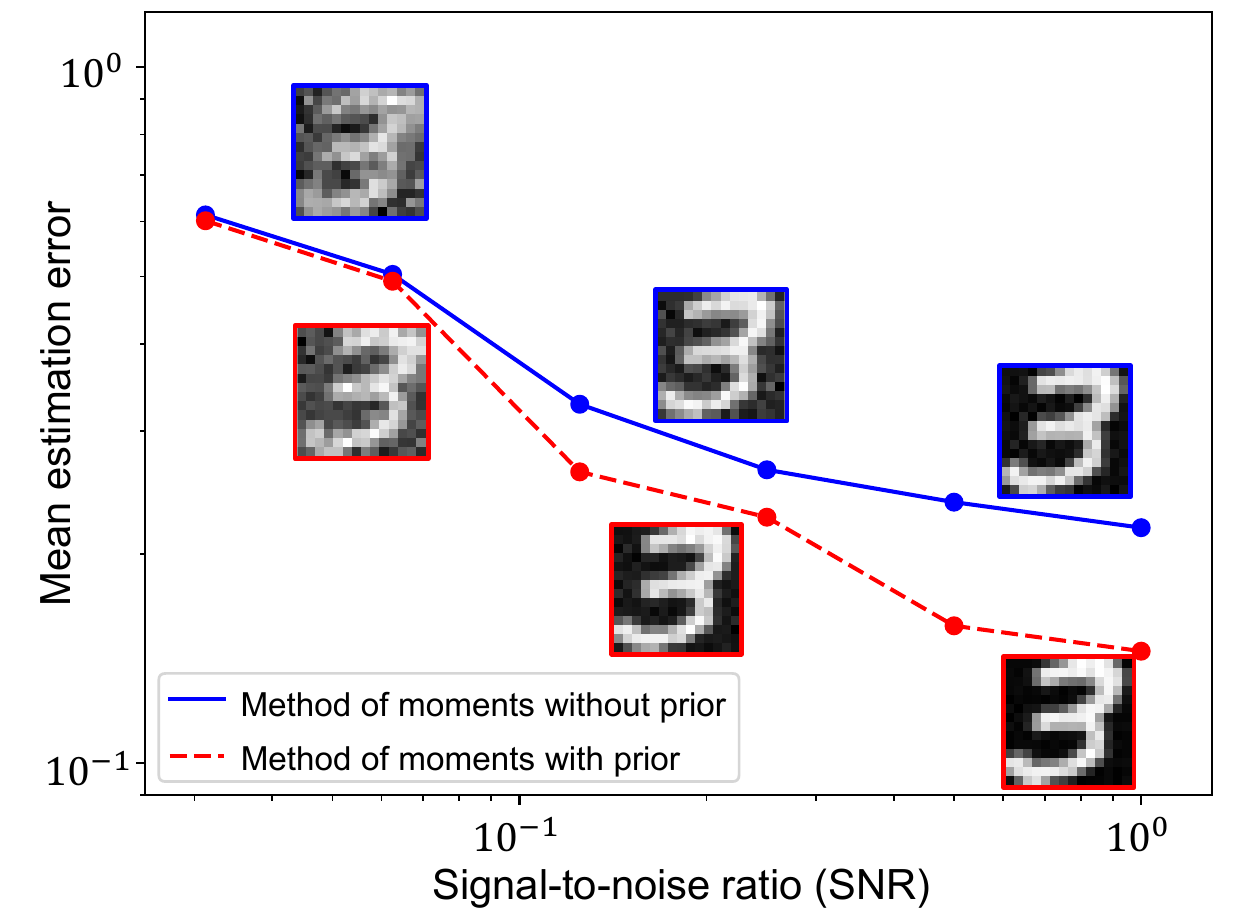}
    \caption{Mean estimation error as a function of SNR for recovering a target image from an MTD measurement using moments up to the third-order, with and without score-based diffusion model prior. Evidently, incorporating the score-based prior consistently improves performance across all SNR levels. To illustrate the visual impact, recovered images for a representative digit are shown at selected SNR values, demonstrating the improvement in reconstruction quality.}
    \label{fig:snr_graph_w_vs_wo_prior_logarithmic}
\end{figure}

In the last dozen years, \emph{data-driven priors} have revolutionized linear inverse problems: plug-and-play methods~\cite{venkatakrishnan2013plug}, GAN priors~\cite{bora2017compressed}, and diffusion models~\cite{kawar2022denoising, chung2022come} have achieved state-of-the-art results in denoising, deblurring, and super-resolution. Yet these advances almost exclusively target \emph{linear} forward models. In this work, we turn our attention to \emph{nonlinear polynomial inverse problems} arising from moment inversion, which are considerably more unstable and remain largely unexplored in the generative modeling literature, and, to the best of our knowledge, have not yet been addressed.
Our results show that such priors can effectively regularize and stabilize these systems, thereby extending the reach of generative models beyond linear operators.

\subsection{The multi-target detection model}
One particularly challenging task involving the inference of signals from low-order moments is the application of the method of moments to the task of multi-target detection (MTD) \cite{bendory2019multi}. MTD entails estimating an image, $x \in \mathbb{R}^{L \times L}$, from a noisy measurement, $y \in \mathbb{R}^{N \times N}$, which contains multiple randomly translated copies of the image. This measurement process is described by: 
\begin{equation}
y[\vec{\ell}] = \sum_{m=1}^M x[\vec{\ell} - \vec{\ell}_m] + \varepsilon[\vec{\ell}], \label{eqn:MTDobservation}
\end{equation}
where $\vec{\ell}_m \in \{L + 1, \ldots, N-L\}^2$ denote \textit{unknown} arbitrary translations, $M$ is the number of occurrences of $x$, and $\varepsilon[\vec{\ell}]$ represents i.i.d. Gaussian noise with zero mean and variance $\sigma^2$. The goal is to estimate the target signal~$x$, while the translations are treated as nuisance variables, and their estimation is not required.
MTD serves as an abstraction for cryo-EM~\cite{henderson1995potential, frank2006three, bendory2020single}, central to efforts in reconstructing 3-D molecular structures from raw data~\cite{bendory2023toward, kreymer2025expectation}. In this work, we focus on the high noise regime and follow~\cite{bendory2019multi, kreymer2022two, bendory2023toward, balanov2025note} in approaching MTD through autocorrelation analysis, a special case of the method of moments.

While autocorrelation analysis has shown some success for MTD, it remains numerically inferior to alternative approaches. In particular, when information is incomplete, as in the super-resolution setting considered later in this paper, the method breaks down entirely. This limitation is especially critical in three dimensions: prior attempts~\cite{bendory2023toward} demonstrated the potential of autocorrelation analysis for 3-D MTD, but also revealed severe numerical challenges. In three dimensions, higher-order moment estimators are markedly more sensitive to noise, and the rapid growth in the number of equations further exacerbates ill-conditioning, making stable recovery increasingly infeasible. These difficulties highlight a critical gap: moment-based methods alone are inadequate in high-dimensional, low-SNR regimes. Bridging this gap requires reconstruction strategies that integrate stronger, data-driven priors to regularize inversion and improve robustness. In this work, we pursue this direction by introducing score-based diffusion models as powerful generative priors for 3-D reconstruction.

\subsection{Score-based diffusion models} 
To overcome the instability of classical moment-based methods, we introduce score-based diffusion models as effective, data-driven priors. Diffusion models~\cite{song2019generative} are probabilistic generative frameworks that progressively corrupt data with Gaussian noise and learn to reverse this process, thereby capturing the underlying distribution of natural signals.

Score-based stochastic differential equations (score-SDE)~\cite{song2020score} is a popular diffusion model that makes use of the Stein score, i.e., the gradient of the log probability density. The key idea behind score-SDE is to perturb the data with a sequence of intensifying Gaussian noise governed by a closed-form stochastic differential equation, and jointly estimate the score functions for all noisy data distributions by training a deep neural network, conditioned on the noise levels. The integration of such priors aims to regularize and solve intrinsically ill-conditioned polynomial systems.

\subsection{Main contribution}
Building upon these ideas, we propose a framework that integrates score-based diffusion priors into the method of moments paradigm to tackle ill-conditioned polynomial systems. We evaluate this approach in the context of MTD through two representative scenarios. First, we study the recovery of the target signal~$x$ via autocorrelation analysis, which gives rise to an ill-conditioned polynomial system. In earlier work~\cite{zabatani2024score}, we demonstrated the potential of score-based priors within the expectation-maximization framework for this setting. Second, we address the super-resolution MTD problem, which is intrinsically ill-posed, as it entails not only recovering the target image but also reconstructing fine-scale features beyond the native resolution of the measurements.

In both cases, the score-based diffusion prior serves to inject data-driven structural knowledge into the recovery process, thereby regularizing the underlying nonlinear polynomial systems and mitigating their sensitivity to noise and underdetermination. More generally, our contribution is a method for solving ill-conditioned systems of nonlinear polynomial equations by combining the statistical consistency of moment-based estimators with the expressive power of modern generative models. We demonstrate this framework on two challenging instances of the MTD problem, showing that it enables robust reconstruction in regimes where traditional techniques struggle. The remainder of the paper develops these ideas in detail: we formalize the problem setting, describe our prior integration strategy, and present empirical evaluations that highlight substantial gains over existing methods.

\section{Mathematical framework}
In this section, we describe a framework for recovering the image $x$ from the MTD observation $y$~\eqref{eqn:MTDobservation}. Our approach combines auto-correlation analysis, which is closely related to the method of moments, with prior information to improve reconstruction quality. 
We begin by defining auto-correlation, and subsequently demonstrate how incorporating priors derived from diffusion models can substantially enhance the accuracy of the recovered signal.

\subsection{Autocorrelation analysis}
We follow~\cite{bendory2019multi} in deriving autocorrelation analysis for MTD. The autocorrelation of order~$q$ of an image~\mbox{$z \in \mathbb{R}^{n \times n}$} is defined as
\begin{equation}
a_z^q[\vec{\ell}_1, \ldots, \vec{\ell}_{q-1}] := \mathbb{E}_z\Big[\frac{1}{n^2} \sum_{\vec{i} \in \mathbb{Z}^2} z[\vec{i}] z[\vec{i} + \vec{\ell}_1] \cdots z[\vec{i} + \vec{\ell}_{q-1}]\Big],
\end{equation}
where~$\vec{\ell}_1, \ldots, \vec{\ell}_{q-1}$ are integer shifts. Indexing out of bounds is zero-padded, i.e.,~$z[\vec{i}] = 0$ out of the range~$\{0, \ldots, {n-1}\} \times \{0, \ldots, {n-1}\}$. As the size of the image grows indefinitely, by the law of large numbers, the empirical autocorrelations of~$z$ almost surely (a.s.) converge to the population autocorrelations of~$z$.
For the MTD problem, our goal is to relate the autocorrelations of the measurement~$y$ to the autocorrelations of the target image~$x$. In practice, we wish to solve the system of equations for $x$
\begin{equation}
\label{eq:mom_equations}
\left\{ a_y^q[\vec{\ell}_1, \ldots, \vec{\ell}_{q-1}] = \gamma a_x^q[\vec{\ell}_1, \ldots, \vec{\ell}_{q-1}] + b_q[\vec{\ell}_1, \ldots, \vec{\ell}_{q-1}]\right\}_{q=1}^3,
\end{equation}
where~$\gamma = \frac{M L^2}{N^2}$ is the known density of the target image occurrences in the measurement, and~$\vec{\ell}_i \in \{0, \ldots, L-1\}^2$. The term~$b_q$ denotes a bias component (see~\cite{bendory2019multi} for details). We adopt the well-separated MTD model, in which each image in the measurement is separated from its neighbors by at least one full image length~$L$. Autocorrelations up to third order are employed, as they are sufficient to uniquely determine a generic signal~$x$ in the MTD model~\cite[Corollary 4.4]{bendory2019multi}.
Solving the system of polynomial equations~\eqref{eq:mom_equations} can be seen as an inverse problem of recovering the signal~$x$ from its autocorrelations.

\subsection{Incorporating priors using score-based diffusion models}
Score-based diffusion models learn the implicit prior of the underlying data distribution by matching the gradient of the log probability density with respect to the data $s(x) \triangleq \nabla_x \log p(x)$.
We approximate the score function using a neural network $s_\theta(x)$, where the network’s parameters $\theta$ are being learned by minimizing the following score matching loss: $\text{loss}(\theta) = \mathbb{E}_{x \sim p(x)}  || s_\theta(x) - \nabla_x \log p(x) ||^2_2.$ For practical considerations and further implementation details, see~\cite{zabatani2024score}.

Leveraging this score-based formulation, we incorporate diffusion priors into the recovery framework by solving a data--prior optimization problem:
\begin{equation}
\label{eq:mom_diffusion}
    \hat{x} = \argmin_x \Big\{ \mathcal{L}_\gamma(x,y)\; -\; \lambda \log p(x) \Big\},
\end{equation}
where the first term enforces consistency with the moment equations in~\eqref{eq:mom_equations}, and the second term incorporates a diffusion prior. The moment loss is defined as
\begin{align}
\label{eq:mom_loss}
    \mathcal{L}_\gamma(x,y) &= \sum_{q=1}^3 \ \sum_{\vec{\ell}_{i=1}^{q-1} \in \{0,\ldots,L-1\}^2}  
    \Big( a_y^q[\vec{\ell}_1,\ldots,\vec{\ell}_{q-1}]  \nonumber \\
    &\quad - \gamma\,a_x^q[\vec{\ell}_1,\ldots,\vec{\ell}_{q-1}] \;-\; b_q[\vec{\ell}_1,\ldots,\vec{\ell}_{q-1}] \Big)^2 .
\end{align}
Here, $p(x)$ denotes the probability density induced by a pre-trained diffusion model prior, and $\lambda>0$ controls its influence. In practice, we assign an adaptive weight to this term for stability. Optimization is performed using Nesterov’s Accelerated Gradient (NAG)~\cite{Nesterov1983AMF}, with the diffusion prior entering only through its score function $\nabla_x \log p(x)$. Since the explicit form of $p(x)$ is unknown, the diffusion model provides an approximation of the score to guide the iterates toward plausible solutions. Algorithm~\ref{alg:imageRecoveryFromAutocorrelations} summarizes the procedure, which applies uniformly to both the standard and super-resolution settings, as elaborated in the next subsection. Full implementation details are provided in Section~\ref{sec:results}.

\begin{algorithm}[t]
\caption{Image recovery from autocorrelations}
\label{alg:imageRecoveryFromAutocorrelations}
\begin{algorithmic}[1]
\Require measurement $y \in \mathbb{R}^{N \times N}$; noise variance $\sigma$; density $\gamma$; score network $S_\theta$ (equals $0$ if no prior); initial guess $x^{(0)} \in \mathbb{R}^{L \times L}$; momentum $\mu$; learning rate $\eta$; score factor $\alpha$; down sampling operator $P$ (replace with $I$ if no super-resolution); sampled index set $\mathcal{M}$ (includes all indices if no super-resolution)
\Ensure An estimate of the target image $x$
\State Calculate the measurement's autocorrelations up to the third-order $a_y^1, a_y^2, a_y^3$
\State Initialize $v^{(0)} \gets 0$
\For{$t = 0, 1, \ldots, T - 1$}
    \State Calculate the gradient term according to~\eqref{eq:mom_loss}:
    \Statex \hspace{\algorithmicindent} $g^{(t)} \triangleq \nabla \mathcal{L}_\gamma\left( P x^{(t)} + \mu P v^{(t)}, y \right)$
    \State Calculate the score term:
    \Statex \hspace{\algorithmicindent} $s^{(t)} \triangleq S_\theta \left( x^{(t)} + \mu v^{(t)} \right)$
    \State Adaptively weight the score term:
    \Statex \hspace{\algorithmicindent} $\tilde{s}^{(t)}[i] \triangleq \begin{cases}
        s^{(t)}[i] \cdot \dfrac{\| g^{(t)} \|}{\| s^{(t)} \|} & i \in \mathcal{M} \\
        s^{(t)}[i] \cdot \alpha & i \notin \mathcal{M}
    \end{cases}$
    \State $v^{(t+1)} \gets \mu v^{(t)} - \eta \left( g^{(t)} - \tilde{s}^{(t)} \right)$
    \State $x^{(t+1)} \gets x^{(t)} + v^{(t+1)}$
\EndFor
\State \Return $x^{(T)}$
\end{algorithmic}
\end{algorithm}

\subsection{Super-resolution}
Super-resolution MTD entails estimating an image,~$x \in \mathbb{R}^{L_{\text{high}} \times L_{\text{high}}}$, from a noisy measurement, $y \in \mathbb{R}^{N \times N}$, which contains multiple randomly translated, down-sampled, copies of the image:
\begin{equation}
\label{eqn:SRobservation}
y[\vec{\ell}] = \sum_{m=1}^M x^{\text{low}}[\vec{\ell} - \vec{\ell}_m] + \varepsilon[\vec{\ell}],
\quad x^{\text{low}}\triangleq Px,
\end{equation}
where $P$ denotes a down-sampling operator that collects $L_{\text{low}} \times L_{\text{low}}$ equally-spaced samples of $x$; $\vec{\ell}_m$ and $\varepsilon[\vec{\ell}]$ are defined as in~\eqref{eqn:MTDobservation}.

The super-resolution MTD problem is inherently ill-posed: without a prior, the signal cannot be recovered at any SNR, as fine details are absent from the measurements. Incorporating a prior makes recovery feasible and improves reconstruction quality with increasing SNR; see~\cite{shani2024denoiser} for an example of such prior incorporation. The procedure is detailed in Algorithm~\ref{alg:imageRecoveryFromAutocorrelations}.

\section{Numerical experiments}
\label{sec:results}
All experiments are based on a modified version of the MNIST dataset~\cite{deng2012mnist}, in which images were cropped to remove marginal padding, thereby avoiding potential ambiguity in digit location within the estimated image.
For all experiments, our measurement consists of 10 sub-measurements, each of size $4000 \times 4000$, with $\gamma = 0.1$, at various noise levels; the effective measurement size is~$N^2 = 16\times 10^7 \text{ pixels}$. For each experiment, we examine 10 target images from the dataset and report the mean estimation error. The estimation error is defined by $E = \frac{\| x^* - \hat{x}\|_\text{F}}{\| x^* \|_\text{F}}$, where $x^*$ denotes the ground truth image, $\hat{x}$ denotes the estimated image, and the subscript~$\text{F}$ denotes the Frobenius norm. We define 
\begin{equation}
    \mathrm{SNR}=\frac{\| x \|_\text{F}^2}{0.25 \pi L^2 \sigma^2}, \label{eqn:snd-def}
\end{equation}
where $0.25 \pi L^2$ is the effective area of the target image in pixels, and $\sigma^2$ is the noise variance of the measurement. The code to reproduce all experiments is publicly available at \href{https://github.com/rabeinhorn/MTD-2D-MoM-diffusion}{github.com/rabeinhorn/MTD-2D-MoM-diffusion}.

\subsection{Estimation from up to the 3rd-order autocorrelation}
The measurement was generated according to \eqref{eqn:MTDobservation}. We follow Algorithm~\ref{alg:imageRecoveryFromAutocorrelations},with parameters~$\mu = 0.993$,~$\eta = 18000$ and $T = 10000$. We use a dataset of MNIST images resized to $L = 14$ as target images. The score network has been trained on
this entire dataset (60,000 samples), excluding the 10 target images that have been tested.
The average estimation errors with and without prior score, as a function of SNR, are presented in Figure~\ref{fig:snr_graph_w_vs_wo_prior_logarithmic}. We show a clear and consistent improvement of the estimation's accuracy when adding the score-based prior to the autocorrelation analysis scheme.

\subsection{Super-resolution}
\begin{figure}[tbp]
    \centering
    \resizebox{0.5\textwidth}{!}{
    \begin{minipage}{\textwidth}
    
        \centering
        \includegraphics[width=0.09\textwidth]{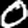}
        \includegraphics[width=0.09\textwidth]{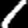}
        \includegraphics[width=0.09\textwidth]{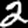}
        \includegraphics[width=0.09\textwidth]{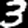}
        \includegraphics[width=0.09\textwidth]{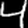}
        \includegraphics[width=0.09\textwidth]{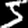}
        \includegraphics[width=0.09\textwidth]{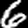}
        \includegraphics[width=0.09\textwidth]{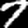}
        \includegraphics[width=0.09\textwidth]{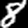}
        \includegraphics[width=0.09\textwidth]{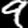}
        \\ [0.5em] {\fontsize{18}{20}\selectfont (a) High resolution images of size $28\times 28$}
    
    \vspace{0.3cm}
    
        \centering
        \includegraphics[width=0.09\textwidth]{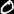}
        \includegraphics[width=0.09\textwidth]{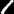}
        \includegraphics[width=0.09\textwidth]{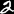}
        \includegraphics[width=0.09\textwidth]{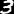}
        \includegraphics[width=0.09\textwidth]{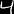}
        \includegraphics[width=0.09\textwidth]{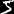}
        \includegraphics[width=0.09\textwidth]{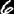}
        \includegraphics[width=0.09\textwidth]{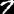}
        \includegraphics[width=0.09\textwidth]{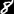}
        \includegraphics[width=0.09\textwidth]{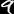}
        \\ [0.5em]{\fontsize{18}{20}\selectfont (b) Low resolution images of size $14\times 14$}
    
    \vspace{0.3cm}
    
        \centering
        \includegraphics[width=0.09\textwidth]{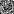}
        \includegraphics[width=0.09\textwidth]{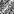}
        \includegraphics[width=0.09\textwidth]{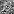}
        \includegraphics[width=0.09\textwidth]{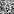}
        \includegraphics[width=0.09\textwidth]{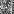}
        \includegraphics[width=0.09\textwidth]{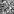}
        \includegraphics[width=0.09\textwidth]{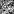}
        \includegraphics[width=0.09\textwidth]{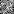}
        \includegraphics[width=0.09\textwidth]{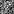}
        \includegraphics[width=0.09\textwidth]{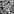}
        \\ [0.5em]{\fontsize{18}{20}\selectfont(c) Low resolution images with noise}

    \vspace{0.3cm}
    
        \centering
        \includegraphics[width=0.09\textwidth]{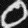}
        \includegraphics[width=0.09\textwidth]{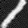}
        \includegraphics[width=0.09\textwidth]{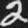}
        \includegraphics[width=0.09\textwidth]{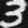}
        \includegraphics[width=0.09\textwidth]{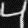}
        \includegraphics[width=0.09\textwidth]{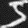}
        \includegraphics[width=0.09\textwidth]{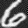}
        \includegraphics[width=0.09\textwidth]{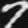}
        \includegraphics[width=0.09\textwidth]{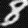}
        \includegraphics[width=0.09\textwidth]{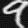}
        \\ [0.5em]{\fontsize{18}{20}\selectfont(d) Estimated high resolution images}

    \end{minipage}
    } 
    \vspace{0.3cm}

\caption{Recovery of all ten digits from the MNIST dataset using the method of moments framework with score-based prior from low resolution target images with $\text{SNR} = 0.5$. The figure shows four rows of images: (a) the original high-resolution images, (b) the downsampled low-resolution images, (c) the low-resolution images with added noise that serve as the input for the recovery algorithm, and (d) the estimated high-resolution images recovered from the noisy inputs.}
    \label{fig:mnist_recovery}
\end{figure}

The measurement was generated according to \eqref{eqn:SRobservation}. We use a dataset of MNIST $28 \times 28$ images as the high-resolution target images, which were down-sampled to create the low-resolution $14 \times 14$ images. The score network has been trained on
the entire dataset of high-resolution MNIST images (60,000 samples), excluding the 10 target images that have been tested. Optimization is done according to Algorithm~\ref{alg:imageRecoveryFromAutocorrelations} with $\mu = 0.993$, $\eta = 18000$, $\alpha = 10^{-10}$ and $T = 5000$.
We show the reconstruction of all digits for a fixed $\mathrm{SNR} = 0.5$ in Figure~\ref{fig:mnist_recovery}, which illustrates the high quality of the recovery achieved.

\section{Discussion and outlook}
In this paper, we introduced a framework that leverages score-based diffusion priors to regularize ill-conditioned polynomial systems arising in moment inversion. Motivated by applications such as the reconstruction of small 3-D molecular structures in cryo-EM~\cite{bendory2023toward}, we focused on the MTD model. Our experiments demonstrated that incorporating a diffusion prior markedly improves recovery from third-order autocorrelations and further enables solutions to the intrinsically ill-posed super-resolution MTD problem.

Looking ahead, several research directions emerge. First, while our approach is empirical, establishing a theoretical foundation for the use of diffusion priors in polynomial inverse problems would offer principled insight into parameter selection and convergence. Second, the framework could be extended to more complex imaging modalities, including cryo-EM and cryo-ET~\cite{bendory2023toward, Chen_2019}, where priors trained on molecular structures may enhance robustness. Third, there is strong potential in designing neural architectures that solve the MTD inverse problem directly, while embedding domain knowledge of autocorrelation analysis for interpretability and efficiency. A related work by the authors is currently exploring a deep convolutional autoencoder to reconstruct a signal directly from its third-order autocorrelations. Finally, our preliminary study of recovery from second-order autocorrelations highlights both the potential and limitations of diffusion priors in even more data-constrained regimes, suggesting a compelling direction for future exploration.

\bibliographystyle{IEEEbib}
\bibliography{refs}

\end{document}